# Confining and channeling sound through coupled resonators

Yun Zhou[1], Prabhakar R. Bandaru[1,2,3], and Daniel F. Sievenpiper[2,3]

*[1]Department of Mechanical Engineering, [2]Program in Materials Science,*

*[3]Department of Electrical Engineering,*

*University of California, San Diego, La Jolla, CA, 92093*

Email: yuz421@eng.ucsd.edu, pbandaru@eng.ucsd.edu, and dsievenpiper@eng.ucsd.edu.

**Abstract:** Confining sound is of significant importance for the manipulation and routing acoustic waves. We propose a Helmholtz resonator (HR) based subwavelength sound channel formed at the interface of two metamaterials, for this purpose. The confinement is quantified through (i) a substantial reduction of the pressure, and (ii) an increase in a specific acoustic impedance (defined by the ratio of the local pressure to the sound velocity) - to a very large value outside the channel. The sound confinement is robust to frequency as well as spatial disorder at the interface, as long as the interface related edge mode is situated within the band gap. A closed acoustic circuit was formed by introducing controlled disorder in the HR units at the corners, indicating the possibility of confining sound to a point.

I. Introduction

The propagation and modulation of sound has been traditionally considered in terms of an acoustic impedance: $Z_{ac}$, through the product of the density ($\rho$) and the velocity ($v_s$) of the medium in which sound propagates. However, it is not easy to understand the confinement of sound, through a traditional *$Z_{ac}$* formulation, as there does not seem to be a reference to which an acoustic impedance may be compared. Moreover, the absence of a magnetic field in acoustic systems does not allow for confinement and related unidirectional/chiral transport [1], without external rotational forces [2–4], implying pseudomagnetic fields are introduced to the system [5]. While acoustic pseudospins [6–13] and valley states [14–18] as related to topological surface states have been proposed to yield directionality, the surface dispersion and associated large velocity favor radiation and consequently a reduced confinement, with unclear robustness to disorder [19]. It may also be expected that wave-based interference phenomena with constructive or destructive interferences could potentially yield regions where sound is focused to be absent or present, respectively, and may be considered for sound confinement [20,21]. However, the intrinsic longitudinal/non-vectorial character of sound propagation is an issue.

An alternate strategy for sound confinement and propagation over a finite distance is to use resonators that are coupled. The flat band [22] related energy dispersions related to local resonators, would permit localization and enhance the possibility for sound confinement. We propose that such *binding* of sound at the subwavelength regime, may be accomplished through Helmholtz resonator (HR)-based arrangements. It has been previously discussed that bands and associated bandgaps could be generated through HRs [23–27], especially at lower energies, while the higher energy bandgaps would be mainly due to Bragg resonances. Patterning HRs onto acoustic topological lattices can render tunability of the Bragg scattering based topological band gaps in the subwavelength range [28–33]. However, band gaps arising from HR related local resonances would not support topological interface states, and the related edge modes or localized states would not generally be subject to a bulk-boundary correspondence. The objective is then to extend the confinement over a number of HR units, the cumulative length over which sound may be considered to be bound. The width of the bands and band gaps could be adjusted through a tuning of the geometrical parameters of the HR. While it has been indicated previously that the coupling of a number of HRs to a waveguide would result in a sound trapping device [34], or negative index acoustic metamaterials [35–37], the related propagation and extent of confinement was not discussed. Moreover, the ability of HR unit-based interfaces to confine sound was not considered, as would be indicated in this work.

## II. Helmholtz resonators constituted unit cell and confined interfacial modes

We consider a unit cell (of lattice constant $a$) comprised of four HRs with differing resonance frequencies ($f$), i.e., $f_4 > f_3 > f_2 > f_1$: **Fig. 1(a)**. The frequencies are normalized by $\frac{c_0}{a}$, where $c_0$ is the velocity of sound in air. As the neck and the cavity of a HR have inductor- and capacitor-like characteristics, respectively, the acoustic inductance for the neck can be written as $L = \rho l_{\text{eff}}/A$, and the acoustic capacitance for the cavity can be expressed as $C = V/\rho c_0^2$, where $\rho$ is density of the air, $l_{\text{eff}}$ is effective length of the neck with end correction, $A$ is the cross sectional area of the neck, and $V$ is the cavity volume. The resonant frequency $f = \frac{1}{2\pi\sqrt{LC}}$ for each HR in the unit cell may be adjusted by tuning the geometry of the neck and the HR cavity [38,39]. Geometrical parameters of HRs are calibrated to yield variation in the $f$ of the unit cell. The corresponding band structure for the unit cell is shown in **Fig. 1(b)**, indicating flat bands and bandgaps characteristic of the HRs (See section A of the **Supplementary Material**). In addition to the length scales, the orientation of a HR would yield a variety of low and higher order couplings, predicated on the interaction of sound dipoles. Consequently, the assembly of the HR composed units would yield rich behavior involving both local and coupled resonances.

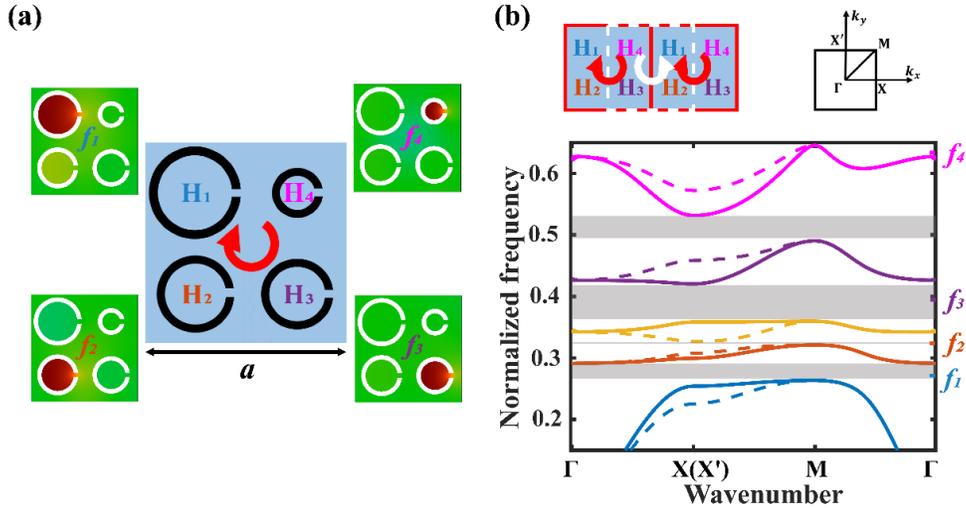

**Figure 1 (a)** Unit cell with Helmholtz resonators arranged in a clockwise order (red arrow) from high resonant frequency to low resonant frequency. The radius of the cavities of $H_1$, $H_2$, $H_3$ and $H_4$ are $\frac{2.2}{12}a$, $\frac{1.8}{12}a$, $\frac{1.6}{12}a$, and $\frac{1}{12}a$, respectively. The length and the width of the necks are $\frac{0.5}{12}a$. $H_1$, $H_2$, $H_3$ and $H_4$ have resonant frequencies of $f_1 < f_2 < f_3 < f_4$. The circular arrow represents the direction of decreasing frequency. **(b)** Band structure of the HR unit cell in (a). Individual resonant frequencies of the resonators are marked in the figure. Solid lines are bands in the direction of $\Gamma X M \Gamma$, while dashed lines are bands in the direction of $\Gamma X' M \Gamma$.

In this paper, we consider a manifestation of the HR unit assembly for possibilities related to robust sound confinement. We study confinement arising from the resonance coupling imbedded in the unit cell. In such an arrangement there is a relative localization of sound in the cell, *e.g.,* as illustrated in the *top inset* to **Fig. 1(b).** However, there is no net overall directionality, as at the interface of two identical and adjacent cells there are oppositely directed HRs, that also can be considered as the unit cell of the bulk. The unit cells are arranged so as to yield cooperative or non-cooperative resonances. The acoustic confinement in the former case is localized over a length scale of the resonator units in the cell. The subsequent channeling can be considered robust locally [40]. Herein, a particular direction is postulated through a specific arrangement of resonators where transport arises from the thermodynamically reasonable flow of energy. The spatial extent of the region along which the sound is confined would also influence the extent of losses, where a larger (/smaller) number of cells would be involved in bounding the acoustic energy over a larger (/smaller) distance of propagation.

We indicate how HR constituted unit cell arrangements can be made, to yield specific bands or states that can be used for acoustic energy confinement. A ribbon supercell of a number of the proposed HR units yields an equivalent band structure in **Fig. 2(a)**, *cf.,* the energy dispersion for a single HR unit cell at the bottom of Fig. 1(b). When an interface is induced, as in *right inset* of **Fig. 2(b)**, through placing two HR constituted bulk structures with unit cells arranged in opposite directions, the hybridization of the energy levels across the interface is expected to yield a multiplicity of two-fold states in the band gap across a range of frequencies, akin to edge states. The related band structure is indicated in **Fig. 2(b),** with the interface (the *right inset*) now in the middle of the ribbon supercell. The clockwise reduction of the frequency (in *red* circular arrows – at the top) and counter-clockwise (in *white* arrows- at the bottom) enhancement, along a line in the ribbon supercell of HR units, together give rise to edge states, originating from the nominal bands [41] as depicted in **Fig. 2(b)**. Subsequently, the energy flow along either direction is enhanced through such edge state modes at the interface, based on resonance coupling and local confinement in the HR unit cells, as previously discussed. However, there is no directionality for the acoustic energy at the interface, over a distance larger than the considered unit cell. This may be seen through considering additionally a unit cell above the one depicted in **Fig. 1(b)**. Here HR units 1 and 4 have on the top as well as the bottom HR units 2 and 3, implying a frequency dependent directionality/gradation *both* at the top and bottom with equivalent energy flow and an overall non-directionality. We noted that replacing half of the unit cells, at the interface, with a sound hard boundary (corresponding to the continuity of velocity along the boundary), yields equivalent effects - see Section H of the **Supplementary Material**. Such sound confinement phenomenon is also not unique to HRs in a square lattice, as similar arrangement of 3 or 6 HRs in a hexagonal lattice would also yield confined edge states, indicated in Section F of the **Supplementary Material**.

Due to local nature of the HR resonances, the edge states do not bridge the bands which helps in the confinement of sound. The edge states may then be analyzed to probe the dispersion and propagation of the confined sound. We will discuss next, the edge mode highlighted in red, in **Fig. 2(b),** which may be construed to be related to maximal confinement, from the large mid-gap ratio [42]. The confinement of sound related to this edge mode is shown through the corresponding pressure field at the bottom: Fig. 2(c), also see *Section B* of the **Supplemental Material**. We have observed, through extensive computational simulations, the sensitivity of the confinement to the orientation of the individual HR units within the unit cell - see *Section C* of the **Supplementary Material**.

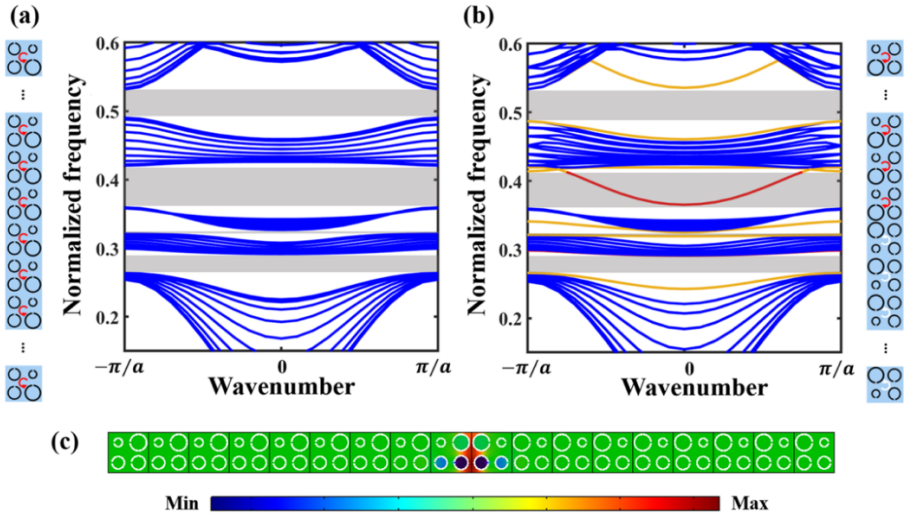

**Figure 2** Band structure of a ribbon supercell composed of **(a)** HR unit cells with resonators arranged in only clockwise order, and of **(b)** HR unit cells arranged in opposite order forming an interface, where additional confined states are found, as highlighted in red and yellow. Periodic

boundary conditions are applied. **(c)** The pressure field corresponding to an eigenmode marked in red in (b).

### III. Local confinement and robustness

For instance, consider a situation where the necks of the HR units are each re-orientated in different directions, so as to face the adjacent HR, as indicated at the top *inset* to **Fig. 3(a).** The edge states associated with the related interface of the modified unit cell is indicated in **Fig. 3(a),** and the related pressure distribution is shown in the bottom *inset* to **Fig. 3(a)**. The necks' orientation won't change the individual resonance frequencies and the band gaps, but influences the cooperative coupling of the HRs and the edge states. From a plot of the phase of the related eigen pressure fields in **Fig. 3(b)**, it was observed that there was an induced phase rotation in the unit cell. It agrees with our previous discussion that the phase rotation direction in the unit cell with half a lattice constant shift has opposite phase rotation direction. The $2\pi$ phase rotation implies a rotational sound energy flux within the unit cell, results in the improved confinement at the interface, compared to the unit cell orientation at the in Figs. 2(b)/2(c). The degree of confinement, related to acoustic energy density, is monitored in the direction perpendicular to the interface. A distance ($L_c$) over which there is a decrease of the pressure amplitude by 3 dB was taken to be the measure of the confinement. At $k = 0.1\frac{\pi}{a}$ (or $\lambda = 20a$), for example, for the interface in **Fig. 2(c),** the $L_c$ was recorded as ~ 0.219 $a$ (or $0.011\lambda$), while for the interfaces depicted in **Fig. 3(a)/(b),** a 26% increase in confinement through a decreased $L_c$ of ~ 0.162 $a$ (or 0.008 $\lambda$) was indicated.

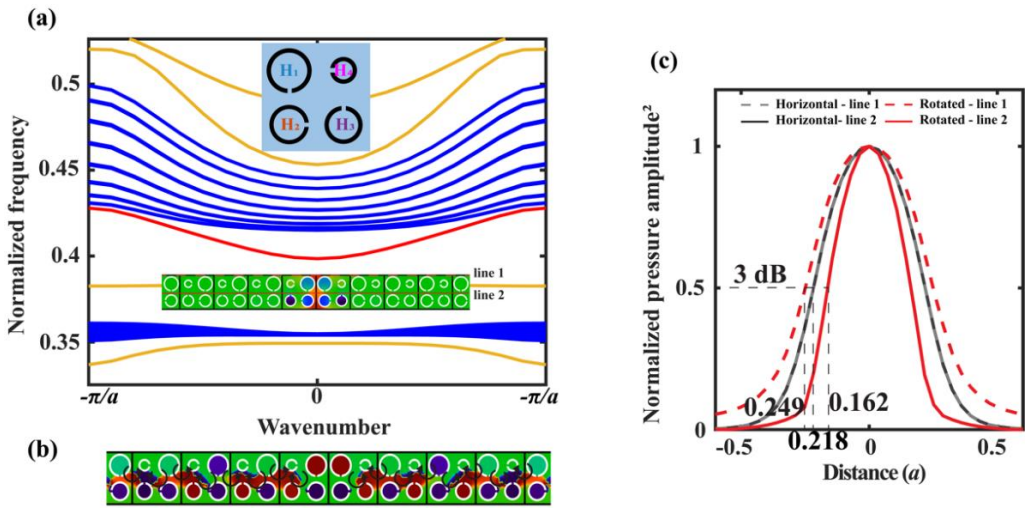

**Figure 3** Tuning of the confinement of the acoustic energy at the interface. **(a)** Band structure of the HR ribbon supercell, where the necks of the HR units are each re-orientated, so as to face the adjacent HR (*top inset*). The *bottom inset* is the eigen pressure field corresponding the *edge state* (in red). **(b)** Phase plot of the eigen pressure field of the edge state mode. **(c)** The normalized pressure $p^2$ vs. the distance perpendicular to the interface, comparing the confinement for an interface configuration, with the HRs in the unit cell all oriented (i) similarly – as in Fig. 2(c), or (ii) differently – as in the inset to (a), for the wavenumber $k = 0.1\frac{\pi}{a}$.

We characterize the confinement of sound through an acoustic impedance ($Z_{ac}$) model, where sound follows a path of minimal impedance. The ratio of the related pressure field to the ensuing local sound velocity ($v_S$): $Z_{ac} = \frac{p}{v_s}$. The related acoustic pressure field for an HR unit cell related interfacial wave guide, excited by a point sound source ($p = p_0 e^{i2\pi ft}$) on the left – indicated by the * with normalized frequency of 0.3848 (wavenumber of $0.44\frac{\pi}{a}$, or 0.115 m$^{-1}$) is plotted in **Fig. 4 (a)**, with the corresponding amplitudes of the pressure (*p*) and velocity (*v*) in **Fig. 4(b)**. In contrast to the rapid decay of the confined sound modes in the perpendicular direction, as in **Fig. 3(c)**, the $p$ and $v_s$ are relatively unattenuated along the propagation direction. Standing wave-like profiles, for the $p$ and $v_s$, through the path of propagation were indicated and subsequently deconvolved so as to yield traveling wave traces (see Section D of the **Supplemental Material**). The standing waves arise from the boundaries (since scattering boundary conditions are used for the simulations) as well as from the periodic and multiple local scattering of the energy from the individual HR units at the interface, rationalizing the sound confinement. The spatial fast Fourier transform (FFT) of the pressure profile in Fig. 4 (c), indicates peaks at wavenumbers of $0 \pm \Delta k, \pm 2 \pm \Delta k, \pm 4 \pm \Delta k, \pm 6 \pm \Delta k$… of $\frac{\pi}{a}$, where $\Delta k$ is the Bloch wave number related to the lattice periodicity, and integer wavenumbers 0, $\pm 2, \pm 4$, …are from the modulation of the unit cell. Specific acoustic impedance of the interface may be estimated through simplifying the local multiple scattering phenomenon to a pair of counter-propagating traveling waves (see Section D of the **Supplemental Material**).

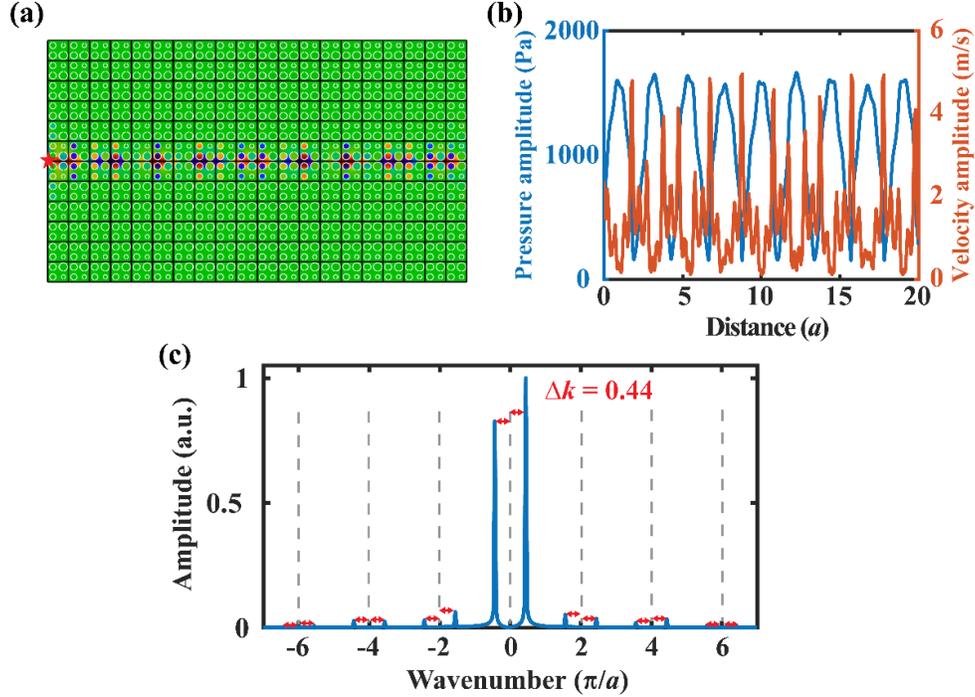

Figure 4 **(a)** Acoustic pressure field at the interface of two metamaterials, constituted from oppositely arranged HR unit cells. **(b)** The magnitude of pressure and velocity along the propagating interface (red dashed line). **(c)** Spatial FFT of the pressure profile along the interface.

We comment further on the essential non-band bridging character of the edge states with the implication of reduced scattering and relative insensitivity to disorder in frequency and spatial arrangement. Such perturbative disorder was introduced in the unit cells at the interface in the ribbon supercell in Fig. 2(b), as shown in **Fig. 5**. Frequency disorder was introduced through enlarging (/reducing) the HR cavity radius, thus decreasing (/increasing) the $f$. Spatial disorder was simulated by displacing the HRs from their original location. While the center column – corresponding to **Fig. 5(b)** indicates the type of disorder, the left column: **Fig. 5(a)** shows the modifications to the band structure, and the right column: **Fig. 5(c)** depicts the resulting acoustic pressure. The sound source is situated at the bottom. The top, middle, and bottom insets of **Fig. 5(b)** illustrate the cases of reduced frequencies, increased frequencies, and spatially displaced resonators: $H_1$ and $H_2$, at the interface. It was seen that fabrication irregularities may be tolerated in the proposed design. From **Fig. 5(a)**, it may be observed that both frequency and spatial disorders tend to push the edge states up or down into the bulk bands. The edge state, and the corresponding energy flow, is robust if the disorder is not so much as to merge the edge band with the bulk bands. **Fig. 5(c)** shows a driven-mode simulation of sound propagation through a HR unit cell constituted waveguide that includes all three types of disorders in **Fig. 5(b)**.

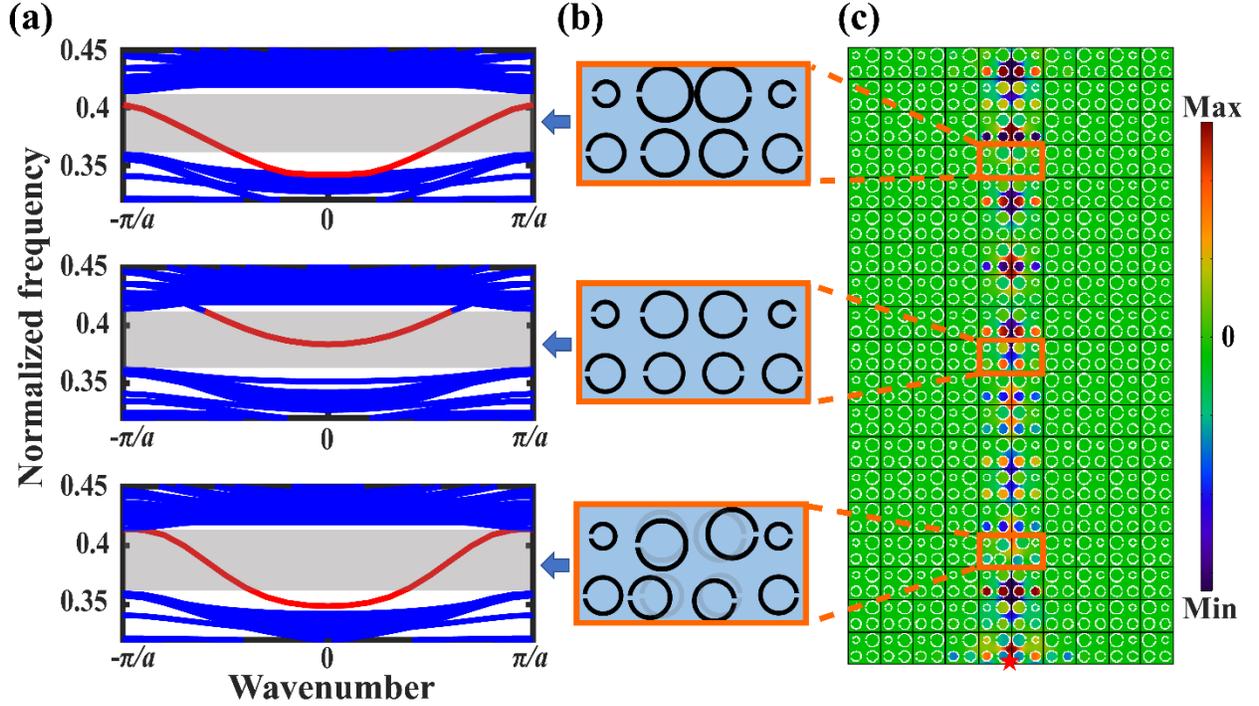

**Figure 5 (a)** Band structure of the ribbon supercell in Fig. 2 (b) with frequency or location disorder at the interface. The corresponding disorders at the interface is shown in **(b)**. Here, <u>Top</u>: interface disorder with reduced frequencies, where the radius of the cavities related to $H_1$, $H_2$ are increased by ~ 10% to $\frac{2.4}{12}a, \frac{2}{12}a$, respectively. <u>Middle</u>: interface disorder with increased frequencies, where the radius of $H_1$, $H_2$ are decreased by ~ 10% to $\frac{2}{12}a, \frac{1.7}{12}a$, respectively. <u>Bottom</u>: interface disorder with displaced $H_1$ and $H_2$. **(c)** The sound pressure corresponding to the indicated frequency and location disorder in (b), along the direction of wave propagation. A point sound source ($p = p_0 e^{i2\pi ft}$ (∗)) is situated at the bottom.

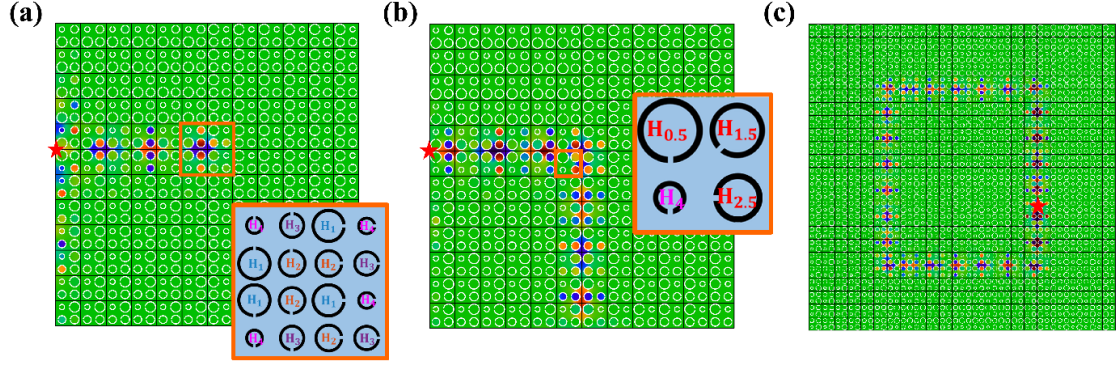

**Figure 6 (a)** A 90° abrupt turn (see closeup of the corner in the *inset*) in the propagation path with significant energy reflection. **(b)** A rearrangement of the corner unit cell by modifying the resonant frequency of the corner HRs (see *inset*), yields smoother turns. **(c)** Acoustic energy confinement in a closed circuit with corner arrangement adapted from **(a)**.

The propagation of sound in **Fig. 4(a)** is over the straight-line path, correspondent to the interface of adjacently placed HR units. Generally, non-collinear sound propagation has been challenging to implement, given that there seems to be no ideal arrangement of the unit cells at the turn regions. At the ends of any line, a localized state is expected which would be either reflected back or dissipated. Alternately, from the viewpoint of impedance matching, the acoustic energy would be not (/totally) reflected if the impedance at the end is matched (/infinitely large) with various degrees of reflection for intermediate cases. We observed that the reflection and transmission could be tailored through HR unit cell configurational changes, *e.g.,* through re-tuning of the frequencies of the resonators at the tuning point, taking advantage of the robustness of the interface modes, discussed previously. For instance, **Fig. 6(a)** indicates an arrangement where almost total stoppage/reflection was seen when the edge mode, propagating over the interface, encounters a 90° sharp turn. An inevitable disorder occurs in the highlighted corner region with detailed arrangement shown in the *inset*, even though the rest of the interface is intact and ordered. The transmission is calculated to be ~ 0.002 (see Section E of **Supplemental Material**). Alternately, a change in the individual HR radius (which tunes the resonator frequency) can be made where considerable transmission of the sound energy around a corner is accomplished: **Fig. 6(b).** The resonance frequencies of the $H_1$, $H_2$ and $H_3$ in the highlighted corner HR unit cell are modified to smaller values, *i.e.*, $f_{0.5}$, $f_{1.5}$ and $f_{2.5}$ ($H_{0.5}$, $H_{1.5}$ and $H_{2.5}$), inducing decreased-frequency disorder to the horizontal interface, and an increased-frequency disorder to the vertical interface in the 90º turn. A dramatic increase in wave transmission was then observed. Consequently, acoustic wave confinement around a closed path is now possible, as shown in **Fig. 6(c).**

However, it is not clear as to what *exactly* the corner configuration ought to be for the continued motion of the sound current. From a consideration of the transmitted power, in terms of the acoustic intensity $I = pv_s$, *before* and *after* the turn, it was noted that while the $p$ is diminished to $0.81p$, the $v_s$ is reduces to $0.69v_s$, which give the total transmission of ~ 56% - see Section E of **the Supplemental Material**) – an increase of ~ 300 compared to the transmission of Fig. 4(d). Perhaps, the magnitude of the transmission may be used as an approximate measure of the suitability of a corner geometric configuration. Future study should focus on the further improvement of the transmission through computational search and related optimization of the corner geometry of the individual HRs.

Moreover, while we have indicated acoustic confinement robust to possible disorders induced by fabrication, in reality, losses are inevitable and need to be taken into consideration. We considered the viscous and thermal boundary-layer induced losses due to acoustic energy dissipation though the narrow necks of the HRs [43–45], and found that the acoustic wave is well confined, *i.e.,* a 3 dB loss is sustained for propagation over 450 unit cells/100 wavelength (see Section G of the **Supplemental Material**).

### IV. Conclusion and future perspective

We have designed a subwavelength acoustic waveguide based on HRs that has great confinement and is robust over lattice disorders. Taking the advantage of robustness of the waveguide, we are able to make sound turn sharp corners. The proposed scheme of HR configuration-based sound confinement can be extended to three-dimensions, where acoustic confinement would now be over an area. While the direction of propagation is based on the orientation of the sound source, the proposed scheme does not allow for unidirectional energy transport. Indeed, given the absence of a magnetic field, Dirac-like points, or nonlinearity in the structure, the propagation of the acoustic waves is yet bidirectional. However, the presence of the HRs in a unit induces a local binding which when added together over several units may be construed as the confinement of sound. Our formulations also allow for an alternate viewpoint of the specific acoustic impedance in terms of the ratio of the driving pressure of acoustic wave propagation to the local acoustic velocity. A closed-path acoustic circuit suggests the possibility of confining sound waves to a very small region, even to a point.

This work was supported by Army Research Office contract W911NF-17-1-0453.